\documentstyle[11pt,pasp3D,twoside,epsf]{article}

\begin{document}

\title{{\sf SAURON}: Integral-field Spectroscopy of Galaxies}

\author{B. W. Miller, M. Bureau, E. Verolme, P. T. de Zeeuw}
\affil{Sterrewacht Leiden, Postbus 9513, 2300 RA Leiden, The Netherlands}

\author{R. Bacon, Y. Copin, E. Emsellem}
\affil{Centre de Recherche Astronomique de Lyon, 9 Avenue Charles
Andr\'{e}, 69561 Saint Genis-Laval Cedex, France}

\author{R. L. Davies, R. F. Peletier, J. R. Allington-Smith}
\affil{Physics Department, University of Durham, South Road, Durham,
DH1 3LE, UK}

\author{C. M. Carollo}
\affil{Dept. of Physics and Astronomy, Johns Hopkins University, 3400
N. Charles St., Baltimore, MD 21218 USA}

\author{G. Monnet}
\affil{European Southern Observatory, Karl-Schwarzschild Str. 2,
D-85748 Garching, Germany}

\begin{abstract}
We present the first results from a new and unique integral-field
spectrograph, {\sf SAURON}.  Based upon the {\tt TIGER} concept, {\sf
SAURON} uses a lens array to obtain two-dimensional spectroscopy with
complete spatial coverage over a field of
$33^{\prime\prime}\times41^{\prime\prime}$ in low-resolution mode
(0\farcs94 lenslets) and of $9^{\prime\prime}\times11^{\prime\prime}$
in high-resolution mode (0\farcs26 lenslets).  The spectra cover the
wavelengths from 4800\AA\ to 5400\AA\ with a spectral resolution of
$\sim\!3$\AA\ ($\sigma\approx75$ km s$^{-1}$). {\sf SAURON} achieved
first light during commissioning on the William Herschel Telescope on
1 February 1999.  We are now commencing a systematic survey of the
velocity dispersions, velocity fields, and line-strength distributions
of nearby ellipticals and spiral bulges.  The wide field of {\sf
SAURON} will be crucial for unraveling complicated velocity
structures.  In combination with available long-slit spectroscopy of
the outer regions of the galaxies, HST spectra of the nuclei, HST
imaging, and dynamical modeling, we will constrain the intrinsic
shapes, mass-to-light ratios, and stellar populations in spheroids.
\end{abstract}

\section{Introduction}

Understanding the formation and evolution of elliptical galaxies and
spiral bulges is complicated by the fact that many, if not all, of
these systems are triaxial or have multiple kinematic components.
Both lead to rich velocity structures that are difficult to map using
long-slit spectroscopy.  In addition, the central few arcseconds of
spheroids are often ``kinematically decoupled'': the inner and outer
regions appear to rotate around different axes (e.g., Illingworth \&
Franx 1989).  This makes {\it two-dimensional} (integral-field)
spectroscopy of stars and gas essential for determining the dynamical
structure of these systems.  Therefore, the galactic dynamics groups
in Lyon, Leiden, and Durham decided to build {\sf SAURON}, a unique
integral-field unit (IFU) with a large field of view and high
throughput that is optimized for studies of the gaseous and stellar
kinematics in galaxies.

\section{The {\sf SAURON} Instrument}

{\sf SAURON} (Spectroscopic Areal Unit for Research on Optical
Nebulae) is a {\tt TIGER}-like integral field spectrograph (Bacon et
al.\ 1995) currently being used at the 4.2-m William Herschel
Telescope of the Isaac Newton Group on La Palma.  The focal plane is
sampled by an array of 1520 square lenses whose projected size on the
sky can be set to 0\farcs94 or 0\farcs26, yielding fields of view of
$33''\times 41''$ or $9''\times11''$, respectively. About 70 dedicated
lenses sample an area offset 1\farcm7 from the main field, to allow
sky subtraction. The spatial coverage is 100\%.  {\sf SAURON}
complements {\tt OASIS} on the CFHT, which allows high spatial
resolution sampling of the nuclear regions (e.g., one can obtain a
4\farcs1$\times$3\farcs3 field with 0\farcs11 sampling for use with the
Adaptive Optics Bonnette).

The grism is optimized for the spectral domain 4800--5400\AA, which
includes important absorption/emission lines such as the Mg triplet,
Fe lines, H$\beta$, [OIII]$\lambda$4959,5007, and [NI]$\lambda$5199. The
instrumental velocity dispersions are $82$~km~s$^{-1}$ (0\farcs94
sampling) or $\sim\!64$~km~s$^{-1}$ (0\farcs26 sampling).

{\sf SAURON} was commissioned on the WHT 1--5 February 1999.  The
instrument performed admirably.  Most of the effort went to focusing
the optics, aligning the CCD, grism, and lenslet array, and adjusting
the baffling to remove stray light and reflections.  To prevent the
spectra from overlapping, the angle between the grism and the lenslet
array was set to 5\fdg19. Flexure was found to be less than 1 pixel in
both the spatial and spectral directions in $\sim\!1800$ seconds.

\section{The {\sf SAURON} Project}

During the project we hope to observe about 80 bright galaxies
covering all Hubble types.  Initially, we are concentrating on E, S0,
and Sa types.  We are restricting the sample to bright ($M_B<-18$)
northern galaxies with $v_{\odot}~<~3000$~km~s$^{-1}$ and available
HST imaging and/or spectroscopy.  This will be a {\em representative}
sample covering the full parameter space of absolute magnitude,
effective radius ($R_{\rm e}$), flattening, rotational support, and
Mg$_b$ index.

We require $S/N > 25$ per resolution element to map the velocity
fields and $S/N>40$ to measure accurate velocity profiles.  Exposure
times of $\sim\!2$ hours per position are needed to reach these levels
at $\mu_B\approx20$ mag arcsec$^{-2}$.  We also aim to map as much of
the inner $1R_{\rm e}$ as possible, using multiple pointings if
necessary.

The first science run was 14--21 February, 1999. Figure 1 shows
preliminary {\sf SAURON} spectra from the centers of NGC~3377 and
NGC~4278.  These spectra were made by summing a strip 3 pixels wide
perpendicular to the dispersion direction.  The sky has been subtracted
but no flat-fielding has been attempted.  {\sf SAURON} can reach
$S/N\approx65$ \AA$^{-1}$ in 1800~seconds in high surface brightness
regions.  Emission and absorption lines from H$\beta$, [OIII], Mg$_b$,
[NI], and FeI are visible.

\begin{figure}
\plotone{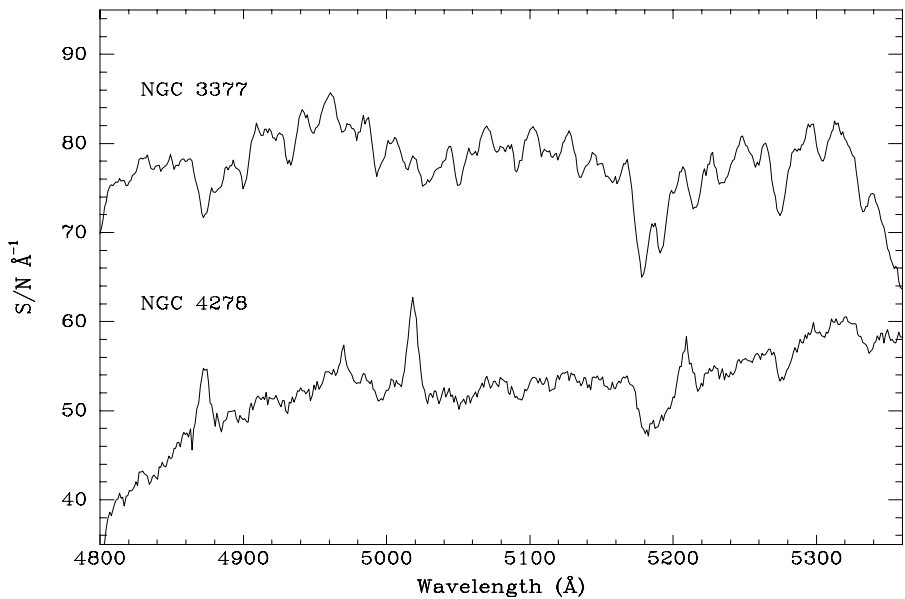}
\caption{Example {\sf SAURON} spectra for NGC 3377 and NGC 4278.  The NGC
3377 spectrum shows absorption lines from H$\beta$, Mg, and Fe.  
NGC 4278 has lines of H$\beta$, [OIII], 
and [NI] in emission as well as Mg and Fe in absorption.}
\end{figure}

The real advantage of {\sf SAURON} is the ability to produce
maps of velocities, velocity dispersions, and line strengths.  Figure
2a shows a preliminary velocity field for NGC~3377.  It reveals a
rotating disk with a maximum velocity of $\sim\!80$~km~s$^{-1}$.  The
velocity dispersion (Figure 2b) is lower in the outer parts of the
disk and then increases towards the nucleus.  Line-strength maps
indicate increasing metallicity towards the center (Davies 1999)

We are currently in the process of refining the data reduction
procedure.  The method will be similar to that used to reduce and
analyze {\tt OASIS} data from the CFHT.
Tight packing of the {\sf SAURON} spectra means that the extraction 
process must be precise.  To account for contamination
from neighboring spectra, we employ an optimized method that
models the distortion of the optics and fits a PSF at each wavelength
of each spectrum.  We are also developing a data reduction pipeline
and a data archive.

\begin{figure}
\plotfiddle{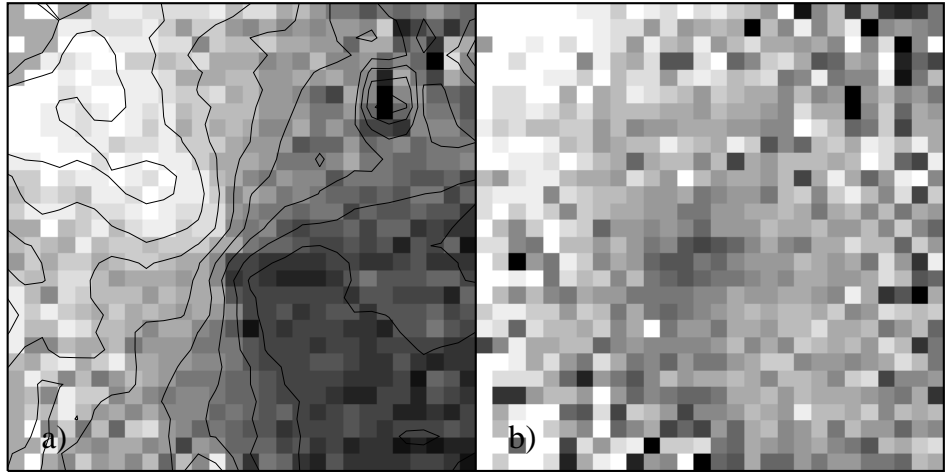}{6cm}{0}{70}{70}{-240}{-350}
\caption{Preliminary {\sf SAURON} maps of NGC 3377. Each panel shows a field
26$''$ on a side. a) Stellar velocity field,
showing evidence for rotation. The contour interval is
20~km~s$^{-1}$. b) Stellar velocity dispersion map, indicating an
increase in $\sigma$ towards the nucleus.}
\end{figure}

The data will be compared with fully general numerical galaxy models
built with Schwarzschild's orbit superposition method.  All
appropriate imaging and spectral data will be used as input for the
models: high resolution HST spectra for constraining the mass of the
central black hole; {\sf SAURON} data cubes for unraveling complicated
velocity structures of the stars and gas within $1R_{\rm e}$;
long-slit spectra for detecting the presence of dark matter halos at
$(1-2)R_{\rm e}$; and HST images and {\sf SAURON} line-strength maps
for estimating the structure and $M/L$ of the starlight.  The galaxy
models make no assumptions about velocity anisotropy and account for
PSF convolution and the finite size of the spatial element (i.e.,
microlens). It has been applied successfully to determine black hole
masses in the E3 galaxy M32 (van der Marel et al.\ 1998) and in
NGC~4342, an E7/S0 in Virgo (Cretton \& van den Bosch 1999).

The output of the {\sf SAURON} project will be a large, uniformly
processed and analyzed sample.  Such a dataset is essential for
understanding the formation and evolution of elliptical galaxies,
spiral bulges, and bars.

\acknowledgments

We would like to thank the Dutch and UK PATTs for allocation of time
on the WHT and the ING staff for their support during the {\sf SAURON}
commissioning and science runs. B.W.M. also thanks the
Leids Kerkhoven Bosscha Fonds for financial support to attend this
conference.

\end{document}